\documentclass[twoside,12pt]{article}
\usepackage{CJK}
\usepackage{indentfirst}
\usepackage{bm}
\usepackage{epsfig}
\usepackage{amsmath}  
\usepackage{xcolor}  
\usepackage[usenames,dvipsnames]{xcolor}

\begin{document}

\begin{CJK*}{UTF8}{gbsn}

\thispagestyle{empty} \vspace*{0.8cm}\hbox
to\textwidth{\vbox{\hfill\huge\sf Commun. Theor. Phys.\hfill}}
\par\noindent\rule[3mm]{\textwidth}{0.2pt}\hspace*{-\textwidth}\noindent
\rule[2.5mm]{\textwidth}{0.2pt}


\begin{center}
\LARGE\bf A brief review of evolutionary game dynamics in the reinforcement learning paradigm$^{*}$
\end{center}

\footnotetext{\hspace*{-.45cm}\footnotesize $^*$Corresponding author, E-mail: zhangjiqiang@nxu.edu.cn.}
\footnotetext{\hspace*{-.45cm}\footnotesize $^\dag$Corresponding author, E-mail: chenl@snnu.edu.cn}

\begin{center}
\rm Guozhong Zheng$^{1,2}$, Xin Ou$^{2}$, Shengfeng Deng$^{2}$, Jiqiang Zhang$^{3, \ast}$, and Li Chen$^{2, \dagger}$
\end{center}

\begin{center}
\begin{footnotesize} \sl
${}^{\rm 1)}$ School of Physical Science and Technology, Inner Mongolia University, Hohhot 010021, PR China \\
${}^{\rm 2)}$ School of Physics and Information Technology, Shaanxi Normal University, Xi'an 710061, P. R. China \\
${}^{\rm 3)}$ School of Physics, Ningxia University, Yinchuan 750021, P. R. China \\
\end{footnotesize}
\end{center}

\begin{center}
\footnotesize (Received XXXX; revised manuscript received XXXX)

\end{center}

\vspace*{2mm}

\begin{center}
\begin{minipage}{15.5cm}
\parindent 20pt\footnotesize
Cooperation, fairness, trust, and resource coordination are cornerstones of modern civilization, yet their emergence remains inadequately explained by the persistent discrepancies between theoretical predictions and behavioral experiments. Part of this gap may arise from the imitation learning paradigm commonly used in prior theoretical models, which assumes individuals merely copy successful neighbors according to predetermined, fixed rules.
This review examines recent advances in evolutionary game dynamics that employ reinforcement learning (RL) as an alternative paradigm. In RL, individuals learn through trial and error and introspectively refine their strategies based on environmental feedback. We begin by introducing key concepts in evolutionary game theory and the two learning paradigms, then synthesize progress in applying RL to elucidate cooperation, trust, fairness, optimal resource coordination, and ecological dynamics. Collectively, these studies indicate that RL offers a promising unified framework for understanding the diverse social and ecological phenomena observed in human and natural systems.
\end{minipage}
\end{center}

\begin{center}
\begin{minipage}{15.5cm}
\begin{minipage}[t]{2.3cm}{\bf Keywords:}\end{minipage}
\begin{minipage}[t]{13.1cm}
reinforcement learning, evolutionary game theory, cooperation, fairness,
trust, resource allocation, biodiversity
\end{minipage}\par\vglue8pt

\end{minipage}
\end{center}

\section{Introduction}
Cooperation~\cite{Nowak2006Five}, trust~\cite{Hardin2002trust}, fairness~\cite{Piketty2014Capital}, and related traits are fundamental to modern society. Understanding their mechanisms is crucial for social stability, human well-being, and fostering a harmonious shared future. From a statistical physics perspective, human society can be viewed as a many-body system, where individuals act as particles and socio-economic activities form the complex interactions among them. Thus, these traits can be viewed as emergence, where we can readily borrow concepts and methods from phase transitions and critical phenomena. A notable example is resource allocation~\cite{arthur1999complexity}, which was formulated as the minority game and is elegantly solved by statistical physicists~\cite{Challet2005Minority}.

Over recent decades, evolutionary game theory~\cite{Smith1973TheLO, Smith_1982Evolution} has advanced our understanding of how such traits emerge. By studying the evolution of prototypical games, theoretical studies have identified key mechanisms underlying these behaviors~\cite{Perc2017statistical}. However, the rise of behavioral economics has revealed persistent inconsistencies between theoretical predictions and experimental observations~\cite{camerer2011behavioral,Arne2010Human}. A significant reason for this gap may be the widespread use of the imitation learning (IL) paradigm in theoretical models~\cite{Nowak1992Evolutionary,Szabo1998Evolutionary}, which assumes individuals simply copy the strategies of more successful neighbors—an assumption often contradicted by experimental evidence~\cite{Angel2018physics}. Real-world decision-making is far more complex than the rigid logic offered by imitation.

To address these inconsistencies, researchers are increasingly adopting a fundamentally different approach: the reinforcement learning (RL) paradigm~\cite{Sutton2018reinforcement}. Within this paradigm, individuals introspectively optimize their strategies through interaction with environments. Crucially, RL emphasizes long-term payoff maximization, contrasting sharply with the short-term, copy-based logic of IL, where imitation may not lead to better outcomes and bypasses individual cognitive reasoning.

This brief review focuses on the reinforcement learning paradigm and examines recent progress in decoding key social traits. In Sec.${2}$, we introduce the framework of evolutionary game theory and compare the two learning paradigms, highlighting the limitations of imitation learning. Sections ${3}$ to ${7}$ present advances in applying RL to cooperation, trust, fairness, resource allocation, and ecological systems, respectively. A summary and future perspectives are provided in Sec. ${8}$.

\section{Fundamentals}
\label{sec:fundamentals}
\subsection{Evolutionary game theory}
Evolutionary game theory (EGT), the core framework of this field, was introduced by John Maynard Smith and George R. Price in their seminal work, \textit{The Logic of Animal Conflict}~\cite{Smith1973TheLO}. By integrating classical game theory with Darwinian evolution, EGT examines how strategies evolve and stabilize within populations, providing a versatile framework for analyzing evolutionary processes in economics, sociology, anthropology, ecology, and beyond.

A central concept in EGT is the evolutionarily stable strategy, analogous to the Nash equilibrium in classical game theory, which describes a strategy resistant to invasion by alternatives under natural selection. In well-mixed populations, the replicator equation~\cite{Taylor1978Evolutionary} describes how strategy frequencies change over time, with each strategy's growth rate proportional to its relative payoff compared to the population average. This formulation captures strategic competition and links evolutionary dynamics directly to the game's payoff structure. EGT has since been widely applied across disciplines, including the study of altruistic behaviors discussed here.

\begin{figure}[tbph]
\centering
\includegraphics[width=1\linewidth]{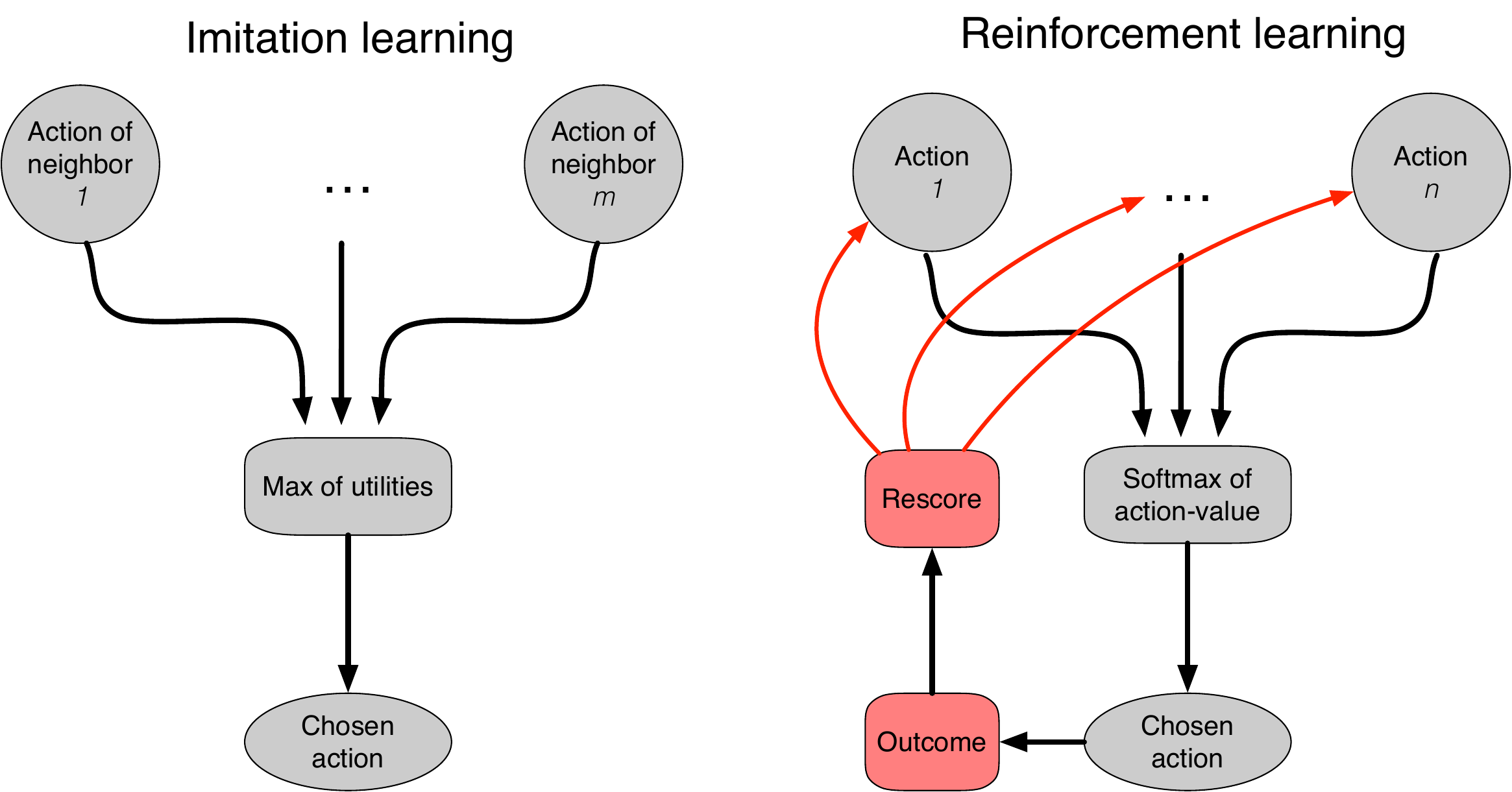}
\caption{Two paradigms for game evolution. In imitation learning, players compare the rewards of their neighbors as the utility and adopt the strategy of the neighbors who have higher utilities. Instead, players with reinforcement learning score different actions and probabilistically choose an action each time based on these scores, which are continuously revised according to the outcome.
}
\label{fig:2paradigms}
\end{figure}

\subsection{The paradigm of imitation learning}
A key component of EGT is the selection process, where better-performing strategies thrive while others diminish. Most theoretical studies model this process using imitation learning (IL), a paradigm in which individuals copy the strategies of more successful peers (Fig.~\ref{fig:2paradigms} Left). Common implementations include the Moran process, the “follow-the-best” rule, and the Fermi updating rule, among others~\cite{Szabo2007Evolutionary}.


{In IL, individuals are relieved of the need to process complex environmental information, making it a simplified representation of human learning—and theoretically, an efficient mechanism for strategy updating. Empirical evidence from behavioral experiments supports this view~\cite{Arne2010Human}, showing that approximately 62\% of strategic choices can be attributed to imitative behavior. This tendency is particularly pronounced in heterogeneous environments characterized by strategic diversity, highlighting both the prevalence and significance of imitation in real-world human decision-making. Furthermore, the concept of imitation translates naturally to ecological contexts, where more adaptive individuals pass on their strategies through genetic inheritance.}

{Nevertheless, IL also exhibits notable limitations. Up to 38\% of strategic changes remain unexplained by imitation alone~\cite{Arne2010Human}, suggesting that human decision-making entails greater complexity than IL typically assumes. This limitation is especially evident in social systems, where individuals do not simply imitate others—particularly when interests are in conflict. Recent behavioral experiments~\cite{Grujic2018comparative} confirm that people often base decisions on others' actions rather than their payoffs, which runs counter to the core assumption of IL. This finding aligns with our intuition that we observe what others do, but not necessarily what they earn. Such discrepancies may help explain why many theoretical predictions fail to align with experimental observations~\cite{Angel2018physics}.}


In essence, imitation learning can be regarded as a simple form of social learning~\cite{Bandura1977social}, where individuals learn from others through observation or instruction in socio-economic activities, which may or may not involve physical practice or direct experience. This means learning occurs by observing the behaviors of others, a manner of ``looking outward".

\subsection{The paradigm of reinforcement learning}
In contrast, reinforcement learning (RL)~\cite{Sutton2018reinforcement} offers a fundamentally different paradigm. Here, individuals learn through direct interaction with their environment, continuously refining their strategies via trial and error (Fig.~\ref{fig:2paradigms} Right). RL emphasizes long-term cumulative reward maximization by balancing experience, immediate payoff, and future expectations. Rooted in psychology and neuroscience~\cite{Daeyeol2012Neural}, RL captures how living organisms learn from experience. 

The RL framework consists of four core elements: a policy (the player's decision rule), a reward (immediate performance feedback), a value function (estimating long-term returns), and the environment. Player-environment interactions typically follow the Markov property, forming a Markov decision process~\cite{puterman2014markov}, which provides the theoretical foundation for RL.

An early RL model is the Bush-Mosteller model~\cite{Bush1955Stochastic}, where players follow the Pavlovian conditioned response, and those winning actions are reinforced. A more influential approach is Q-learning~\cite{Watkins1989learning}, a model-free, value-based algorithm that learns a Q-value function for state-action pairs to guide decision-making. Using an off-policy update rule, it efficiently explores the optimal policy in discrete action spaces.

In Q-learning, a Q-table stores values $Q(s,a)$, representing the expected cumulative reward for taking action \( a \) in state \( s \). The larger the value $Q(s,a)$ within state \( s \), the more preferred action \( a \) is. In the $\epsilon$-greedy Q-learning, players independently choose a random action with probability $\epsilon$; otherwise, they select the action with the largest Q-value within the row corresponding to their states. 
Afterwards, the Q-values are updated at the end of each round via the Bellman equation~\cite{Sutton2018reinforcement}:
\begin{equation}
\begin{split}
Q\left(s_t,a_t\right)\leftarrow(1\!-\!\alpha)Q\left(s_t,a_t\right) \!+\! \alpha \left[ \Pi_{t+1} \!+\! \gamma \max_{a'} Q\left(s_{t+1}, a'\right) \right].
\end{split}
\end{equation}
Here, \( \alpha \) is the learning rate, which determines how much the old experience $Q(s_t,a_t)$ is removed -- the smaller the value, the more experience is retained. \(\Pi_{t+1}\) is the immediate reward obtained by taking action \( a \) in state \( s \) at time $t$.  \( \gamma \) is the discount factor determining the impact of the optimal action in the next step $t+1$ that one can expect. A larger \( \gamma \) means the player values more rewards in the future, having a long-term vision. New values integrate the experience in the past, the reward in the current step, and guidance from the future. This unambiguous interpretation makes Q-learning a mainstream choice in RL applications.


{While actions are often fixed in game-theoretic contexts, the definition of the state is highly flexible and central to the functioning of reinforcement learning. The power of RL lies in its ability to condition actions on environmental information, with the state serving as the representation of that environment—it determines what information the individual perceives. As expected, the design of the state space significantly influences the effectiveness of RL. On one hand, appropriate state representations that provide sufficient environmental information are essential for good performance; on the other hand, including too much information may exceed individuals' memory and learning capacities—capacities that are often assumed to be unlimited in theoretical models but are bounded in reality. In the following sections, we explore a range of state designs, from simple self-regarding setups to other-regarding configurations that incorporate neighbor states, from symmetric to asymmetric information structures, and from precise to fuzzy information representations. These varying designs not only affect individuals' learning efficiency and the resulting behaviors but also shape the model's capacity to capture real-world complexity.}

Other value-based RL variants include SARSA (an on-policy alternative) and deep Q-learning, which uses neural networks to handle large and continuous state spaces. 
Another category of RL is policy-based methods~\cite{Williams1992Simple,NIPS1999_464d828b}, where agents directly learn what actions to take without scoring them. Actor-critic algorithms integrate both value-based and policy-based methods, achieving a balance between learning efficiency and stability. {Moreover, some more advanced reinforcement learning techniques, such as deep RL methods~\cite{vincent2018introduction}, the multi-agent RL~\cite{albrecht2024multi}, and inverse RL~\cite{arora2021survey}, among others~\cite{wang2022modelling}, have shown great potential in handling complex game-theoretic scenarios and also deserve further exploration and attention.}

{In short, IL and RL represent two distinct learning logics suited to different research contexts. IL focuses on leveraging existing experience through observation, corresponding to a ``follow-the-crowd" social heuristic. In contrast, RL adopts an introspective, experience-driven approach, allowing agents to develop strategies through their own interactions with the environment—capturing a process of introspective exploration characterized as ``learning from consequences." The choice between these two paradigms ultimately depends on the specific research context and the questions under investigation.}

\section{Cooperation}\label{sec:cooperation}
Cooperation is widespread and essential in both human societies and natural systems, and the mechanisms underlying its emergence have been extensively studied~\cite{Nowak2006Five}. In the imitation learning (IL) paradigm~\cite{smith1982evolution}, a key requirement is that individuals must share their strategies and payoffs during the evolutionary process -- an assumption that is often neither feasible nor realistic. Because of this limitation, game-theoretic predictions under IL frequently fail to align with behavioral experiments~\cite{Angel2018physics}, making the emergence of cooperation a persistent challenge. Recently, reinforcement learning (RL) has provided a fundamentally different approach to addressing this problem, emerging as a promising paradigm for deciphering the origins of cooperation~\cite{Xie2026reinforcement}.

\subsection{The pairwise game}
The prisoner’s dilemma game (PDG) is a classic pairwise model used to study cooperation. In this two-player, two-action ($2\times2$) game, each player can either cooperate or defect. Mutual cooperation yields both players a reward $R$, whereas mutual defection results in a punishment $P$. If one cooperates while the other defects, the defector receives a temptation payoff $T$, and the cooperator receives the sucker’s payoff $S$. Payoffs in the PDG satisfy $T > R > P \geq S$ and $2R > T + {S}$. In its standard form, the payoff matrix can be expressed as
\begin{align}\label{eq:PDG_payoff}
	\Pi = 
	\begin{bmatrix}
		R & S  \\
		T & P 
	\end{bmatrix} =
		\begin{bmatrix}
		1 & -b  \\
		1+b & 0 
	\end{bmatrix},
\end{align}
where the temptation factor $b$ quantifies the conflict between individual and collective interests. The dilemma arises because defection yields a higher individual payoff regardless of the opponent's choice, even though mutual cooperation maximizes collective welfare.

\begin{figure}[tbp!]
\centering
\includegraphics[width=1.0\textwidth]{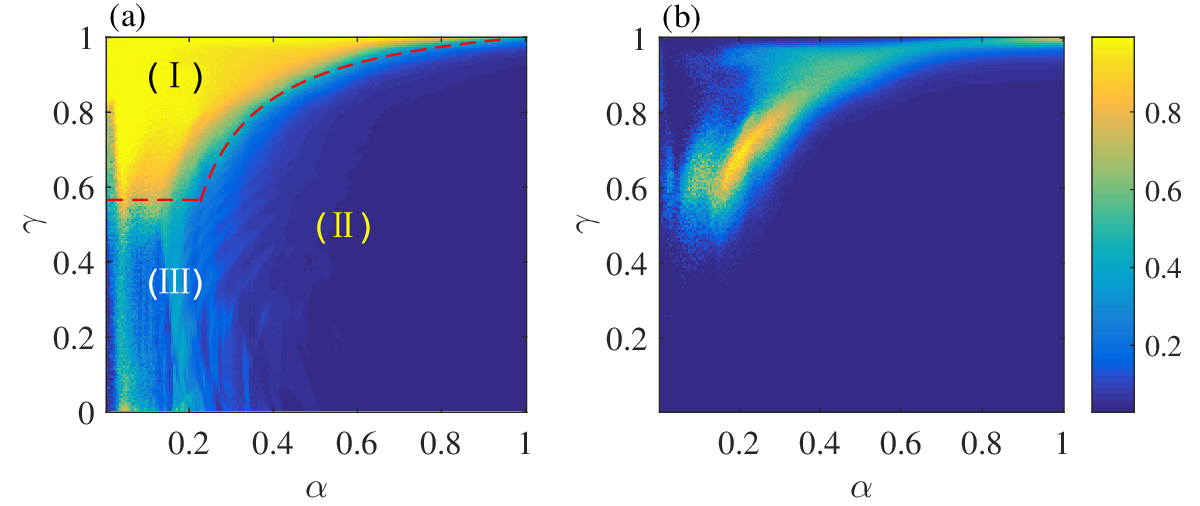}
\caption{Emergence of cooperation in the prisoner’s dilemma game. 
(a) The phase diagram of cooperation level within the space of learning parameters $(\alpha, \gamma)$ when two players play the game, which can be divided into three regions: high cooperation (I), full defection (II), and low cooperation (III).  
(b) The corresponding reward difference between two players, which is visible around the boundaries between Regions I--II and I--III.
$b = 0.2$ in Eq.~\eqref{eq:PDG_payoff} and $\epsilon = 0.01$
(Adapted from~\cite{ding2023emergence}).
} 
\label{fig:phase_diagram} 
\end{figure}

Ref.~\cite{ding2023emergence} represents an early application of RL to explain cooperation in the PDG, focusing on the fundamental dynamics. A key finding is that cooperation emerges when players both value past experience and adopt a long-term perspective (Region I in Fig.~\ref{fig:phase_diagram}(a)). Cooperation also persists at a moderate level when players learn from history, even if they are short-sighted (Region III). Notably, high inequality in payoffs occurs at the boundary between Regions I and II, while rewards are nearly equal elsewhere [see Fig.~\ref{fig:phase_diagram}(b)]. Mechanistically, the study identified that players adopt a win-stay-lose-shift strategy to sustain cooperation and converge to some coordinated optimal modes. When players ignore experience or become myopic, these modes destabilize due to unpredictable opponent behavior, and defection prevails.

{A critical aspect of RL in modeling cooperation is the design of the state -- i.e., what information players perceive. Early studies often used a self-regarding setup, where the state only included the player's own previous action. This, however, provides insufficient information to capture their surroundings. Conversely, overly detailed state representations can be infeasible and may contain redundant information, since decisions often depend on only a few pieces of key cues. Ref.~\cite{zheng2025evolution} illustrates how information perception influences outcomes in the PDG: when two players operate under different information scenarios, the resulting evolutionary dynamics vary significantly, with particularly rich behaviors emerging under asymmetric perception.}

Many studies have extended the two-player PDG to multi-agent settings, such as on 2D lattices, to investigate cooperation at the population level~\cite{ding2019q,lee2025enhancing}. Within RL, classical mechanisms from imitation learning have been revisited, including direct reciprocity~\cite{jia2021local}, indirect reciprocity~\cite{zhao2024emergence}, spatial reciprocity~\cite{wang2025spatial}, adaptive migration~\cite{fang2025evolution}, reputation~\cite{zhang2025cooperation}, and preferential selection~\cite{bai2024preferential}. Regarding spatial reciprocity, Wang et al.~\cite{wang2025spatial} demonstrated that it emerges only when players interact and learn within overlapping local neighborhoods, highlighting the importance of coupling local interaction with local strategy learning for sustaining cooperation under RL. Furthermore, introducing third parties -- such as leaders~\cite{ding2019q}, exporters~\cite{you2023cooperative}, or loners~\cite{huang2025promoting} -- has been shown to effectively promote cooperation under RL, consistent with earlier IL findings.

Beyond these well-known factors, RL-based studies of the PDG have revealed novel phenomena~\cite{zhang2012universal,wang2022levy,wang2023reinforcement}. Zhang et al.~\cite{zhang2012universal} found that moderate greediness optimally promotes cooperation, a result robust across network types. In Ref.~\cite{jia2021local}, global players who consider neighborhood-wide stimuli foster stronger conditional cooperation via direct reciprocity, thereby helping to sustain cooperation. Wang et al.~\cite{wang2022levy} showed that L{\'e}vy noise in payoffs enhances cooperation by creating a Q-value advantage for cooperative actions -- an effect absent under Gaussian or no noise. In a DQN-based updating model, Ref.~\cite{wang2023reinforcement} reported that increasing the discount factor expands cooperative clusters until full cooperation is achieved, while the temptation factor $b$ has little influence. 
{Recently, Su et al.~\cite{su2025multi} employed multi-agent reinforcement learning to discover a novel strategy termed the ``memory-two bilateral reciprocity" (MTBR) strategy. This strategy not only outperforms most known strategies in pairwise interactions but also dominates in evolving populations, promoting higher levels of cooperation and social welfare. This remarkable performance has been validated through both simulations and mathematical analysis, highlighting that multi-agent reinforcement learning not only serves as a strategy update mechanism but also demonstrates significant potential as a strategy discovery tool. Interestingly, the overall cooperation is also enhanced -- a catalytic-like effect was also discussed in Ref.~\cite{sheng2024catalytic}.}

\begin{figure}[tbph]
\centering
\includegraphics[width=1\linewidth]{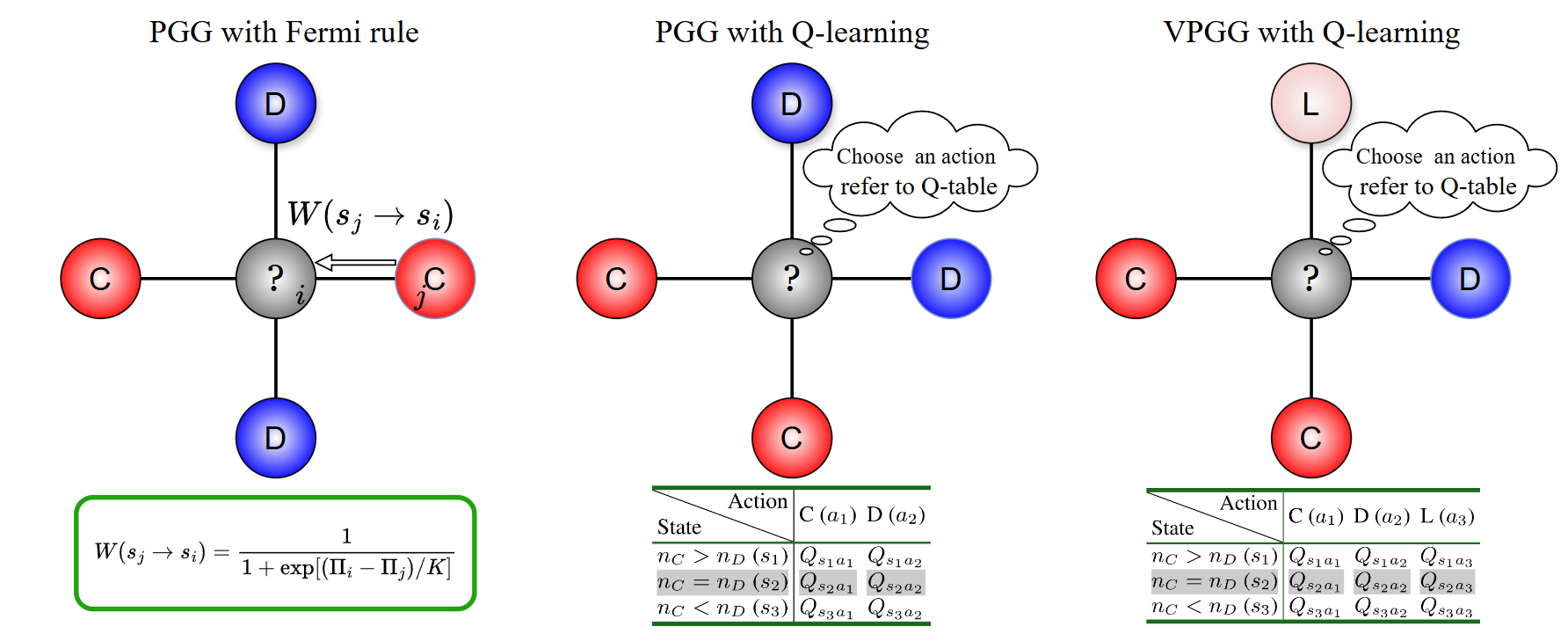}
\caption{Schematics of three model setups. (Left) Public goods game (PGG) with a Fermi-function update rule. (Middle) PGG with Q-learning. (Right) Volunteer public goods game (VPGG) with Q-learning. The focal individuals are indicated by gray and are surrounded by neighbors that could be cooperators (red), defectors (blue), and loners (pink). Below each panel are the corresponding strategy-update components: the Fermi function (left) and the Q-tables (middle and right).
(Adapted from~\cite{zheng2024evolution}).
}
\label{fig:ZhengPGG}
\end{figure}

\subsection{The multi-player game}
Beyond the pairwise game, the public goods game (PGG) serves as a paradigmatic model of multi-player cooperation, where an arbitrary number of participants are allowed in a single game. In a typical PGG, $n$ players may each contribute an amount $a\in[0,1]$ to a common pool. The total pool is then multiplied by a synergy factor $r$ and is divided evenly among all players. The payoff for player $i$ is
\begin{align}\label{eq:payoff_PGG}
\Pi_i &= \frac{r}{n}\sum\limits_{j=1}^{n}a_{j}-a_{i} = \hat{r}\sum_{j=1}^{n}a_{j}-a_{i},
\end{align}
where $a_{i}$ denotes player $i$'s contribution and $\hat{r}=\frac{r}{n}$ is the reduced synergy factor. If $\hat{r} >1$, the contribution $a_i=1$ is expected. However, for $\hat{r} < 1$, the dominant strategy for a rational player is to contribute nothing, i.e., $a = 0$. In this case, contributing while others free-ride increases group benefit at a personal cost, leading to widespread free-riding and the tragedy of the commons~\cite{hardin1968tragedy}. In many practices, $a$ is chosen to be discrete within $\{0,1\}$, corresponding to defection and cooperation, respectively.

Recent studies have applied RL to explore cooperation in PGGs, offering new mechanistic insights~\cite{wang2023synergistic,zhang2024exploring,zou2024incorporating,zheng2024evolution,li2025cooperation}. As with PDG research, some works examine whether cooperation mechanisms known from imitation learning also operate under RL, including reward incentives~\cite{wang2023synergistic}, voluntary participation~\cite{zhang2024exploring}, and reputation systems~\cite{zou2024incorporating,kang2025neighbor}. For example, adaptive reward schemes integrated with self-regarding Q-learning can significantly raise cooperation levels~\cite{wang2023synergistic}. Introducing third-party ``loners" leads to stable cooperation at high synergy factors, while defector density exhibits a non-monotonic dependence on the gain factor~\cite{zhang2024exploring}. The PGG is also studied on higher-order networks~\cite{xu2024reinforcement}, where engagement is determined by Q-learning, and active players utilize social learning to act. Reputation mechanisms on hypergraphs also promote cooperation, with learning parameters being systematically examined~\cite{zou2024incorporating}. Other factors such as neighbor influence~\cite{kang2025neighbor} and conformity~\cite{zhang2025combined} are also investigated. Altogether, these studies further enrich our understanding of PGG dynamics under RL.

Most of these studies, however, adopt a self-regarding state setup, where agents base their decisions solely on their own past actions — contrary to real-world action logic, where individuals also observe their surroundings and respond to their neighbors. Ref.~\cite{zheng2024evolution} emphasizes the importance of social information in RL, comparing three models: traditional IL-based PGG, Q-learning-based PGG, and a voluntary PGG (VPGG) with Q-learning that includes a “loner” strategy on a lattice (Fig.~\ref{fig:ZhengPGG}). Using Fermi updating for IL as a baseline and coarse-graining surroundings by comparing cooperator/defector counts, they show that Q-learning substantially reduces the critical synergy factor $\hat{r}$ needed for cooperation, with VPGG lowering it further. Loners suppress the spread of defectors, although cooperation exhibits non-monotonic dependence on parameters. Again, cooperation is strongest when players value past experience and maintain a long-term perspective.

Ref.~\cite{li2025cooperation} further demonstrates the value of other-regarding information on spatial hypergraphs. Interestingly, two abrupt transitions in cooperation emerge as $\hat{r}$ varies, separating three regimes: no cooperation, moderate cooperation, and high cooperation. Spatial analysis reveals a chessboard-like pattern that promotes the first transition but hinders the second. Theoretical analysis of the first transition shows that far-sighted players with low exploration rates are more likely to reciprocate cooperation, thereby facilitating its emergence.

Notably, the IL and RL paradigms are not mutually exclusive. Human behavior often switches between different decision logics depending on context, a diversity observed in experiments~\cite{Traulsen2010Human}. Several studies capture this by allowing mixed updating rules, yielding rich evolutionary dynamics -- for example, combining social learning with self-learning~\cite{Han2022hybrid}, Q-learning with Fermi rule~\cite{zhang2025layered,yang2025evolution}, Fermi rule with tit-for-tat~\cite{ma2023emergence}, Q-learning with tit-for-tat~\cite{sheng2024catalytic}, and hard with soft conditional cooperators~\cite{zhao2025evolution}, etc. These works frequently identify an optimal mixing ratio at which cooperation exceeds levels achieved under either pure rule set.

\begin{figure*}[ht]
 \centering
\includegraphics[width=0.7\linewidth]{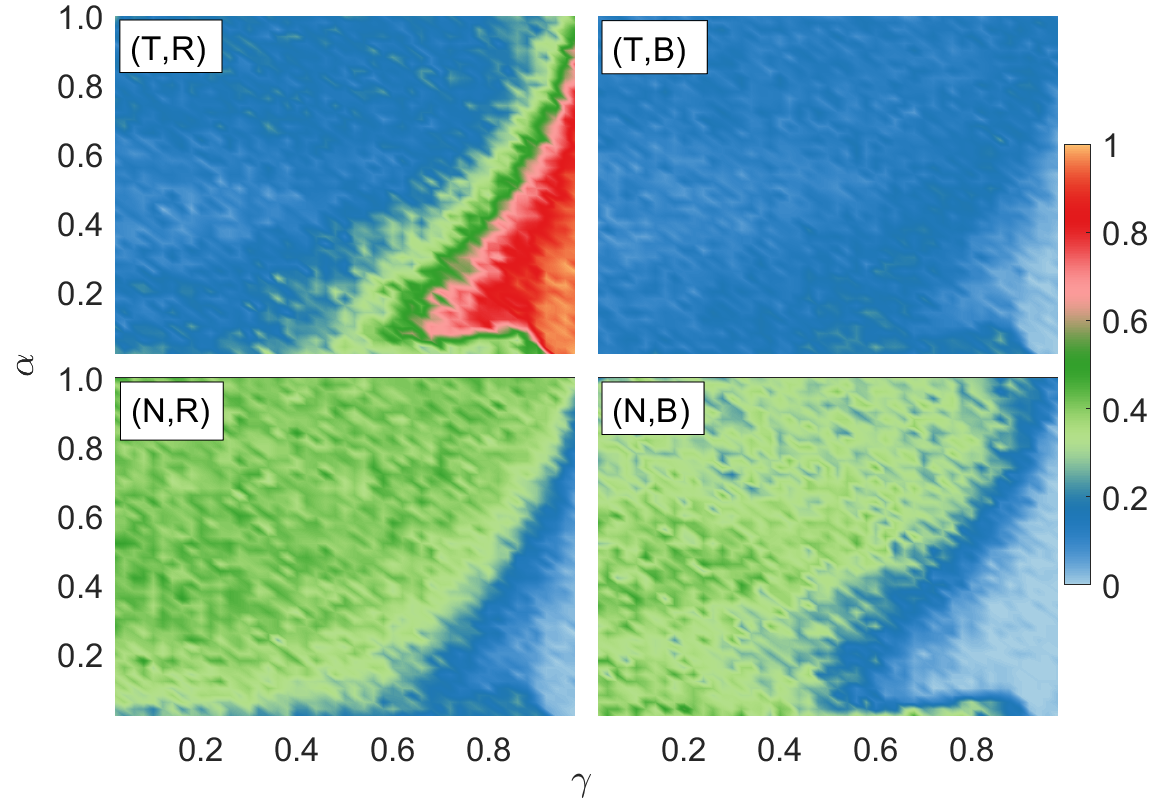}
\caption{Emergence of trust. Fractions of four strategies in the trust game within the parameter space $(\gamma, \alpha)\in (0,1)$. For instance, TB denotes a player who trusts as a trustor but betrays as a trustee. High levels of trust (TR) emerge at large  $\gamma$ and small $\alpha$ (bottom-right corner), indicating that appreciating both historical experience and long-term vision promotes trust and trustworthiness.
(Adapted from~\cite{Zheng2024Decoding}).
}
 \label{fig:trust}
 \end{figure*}

\section{Trust}\label{sec:trust}
Trust, as a core element of human civilization~\cite{Hardin2002trust}, is often regarded as the ``lubricant of the social system" and plays an irreplaceable role in facilitating cooperation and fostering social coordination~\cite{Arrow1974limits}. At the individual level, trust helps establish healthy interpersonal relationships and social networks, promoting collaborative partnerships and the formation of friendships. At the society level, it enhances well-being and quality of life. From an institutional perspective, trust between the government and the public not only strengthens citizens' understanding and identification with the political system but also ensures the effective formulation and implementation of public policies, thereby contributing to social stability and development~\cite{Zak2001trust,Algan2013Trust}.

Research has often employed the trust game~\cite{Berg1995Trust} to study the evolution of trust, in which a trustor decides whether to invest part of their endowment in a trustee with trust (T) or keep it (No trust). If they keep it, the game ends. If they invest, the amount is multiplied, and the trustee then chooses either to reciprocate (R) by returning some money or to betray (B) by keeping everything. According to the assumption of \emph{Homo economicus} in the classic economics -- that individuals are rational and self-interested, aiming to maximize their own payoff --- the trustee should always betray, and the trustor, anticipating this, should never invest. Trust, therefore, would not be expected to arise.
However, behavioral experiments using the trust game~\cite{Berg1995Trust} reveal a fundamental contradiction: trust and trustworthiness are widely observed in humans. On average, trustors invest about 50\% of their endowment, and trustees return roughly 37\% of the gains~\cite{Johnson2011Trust}, demonstrating the pervasiveness of trust in human interactions.

To resolve this discrepancy, prior game-theoretic studies have incorporated factors such as reputation~\cite{Bravo2008The}, population structure~\cite{Wang2024Evolution,Guo2023Evolution}, migration mechanisms~\cite{Zhu2024Payoff}, third-party deposit systems~\cite{Guo2024Evolution}, and incentive schemes~\cite{Liu2025Evolution} into models, showing that these can trigger the emergence of trust. Notably, these works operate within the IL paradigm, where individuals replicate successful strategies to explain the spread of trust in populations~\cite{Kumar2020The}.

Recently,  studies have shifted toward the RL paradigm and shown that endogenous factors alone are sufficient to explain the emergence of trust; no exogenous factors are needed. Ref.~\cite{Zheng2024Decoding} adopts the Q-learning algorithm, focusing on the compromise between short-term self-interest and long-term trust benefits in the trust game. The study mainly discussed the two-player scenarios, where the two players play the role of trustor and trustee in turn. Accordingly, each player is associated with two Q-tables to guide the decision-making for the two roles, respectively. The study revealed that when individuals appreciate both their historical experience (a small learning rate $\alpha$) and the returns in the future (a large discount factor $\gamma$), a high level of trust emerges naturally.
As seen in Fig.~\ref{fig:trust}, the proportion of the trust strategy (TR) peaks in the bottom-right corner, corresponding to low $\alpha$ (value historic experience) and high $\gamma$ (long-term perspective). Q-table analysis reveals a shift in preference from short-term gain to long-term reciprocity, illustrating how trust stabilizes over time. These findings remain robust when extended to a one-dimensional lattice population.

More recently, Ref.~\cite{Zhu2025Q} expanded this research to a spatial lattice by combining Q-learning with second-order social norms, exploring the synergy between reputation and learning in trust evolution. They show that Q-learning yields a richer set of steady-state strategies than the traditional Fermi rule, and even under high dilemma intensity, group wealth can improve. Ref.~\cite{hu2026higher} further extends the model to higher-order networks with reputation mechanisms, which significantly enhance trust and collective wealth accumulation.

Though trust theory in the RL paradigm is still in its infancy, its logic aligns with existing behavioral evidence. For example, Ref.~\cite{Jim2004The} finds that future-oriented concern is essential for sustaining trust in repeated trust games: experienced subjects show much lower trust in a definite round of repeated games than in indefinitely repeated ones, where no future rewards are expected. This forward-looking motivation -- beyond the immediate payoff focus of IL -- is naturally captured within the RL framework.

 \begin{figure}[ht]
 \centering
\includegraphics[width=0.9\linewidth]{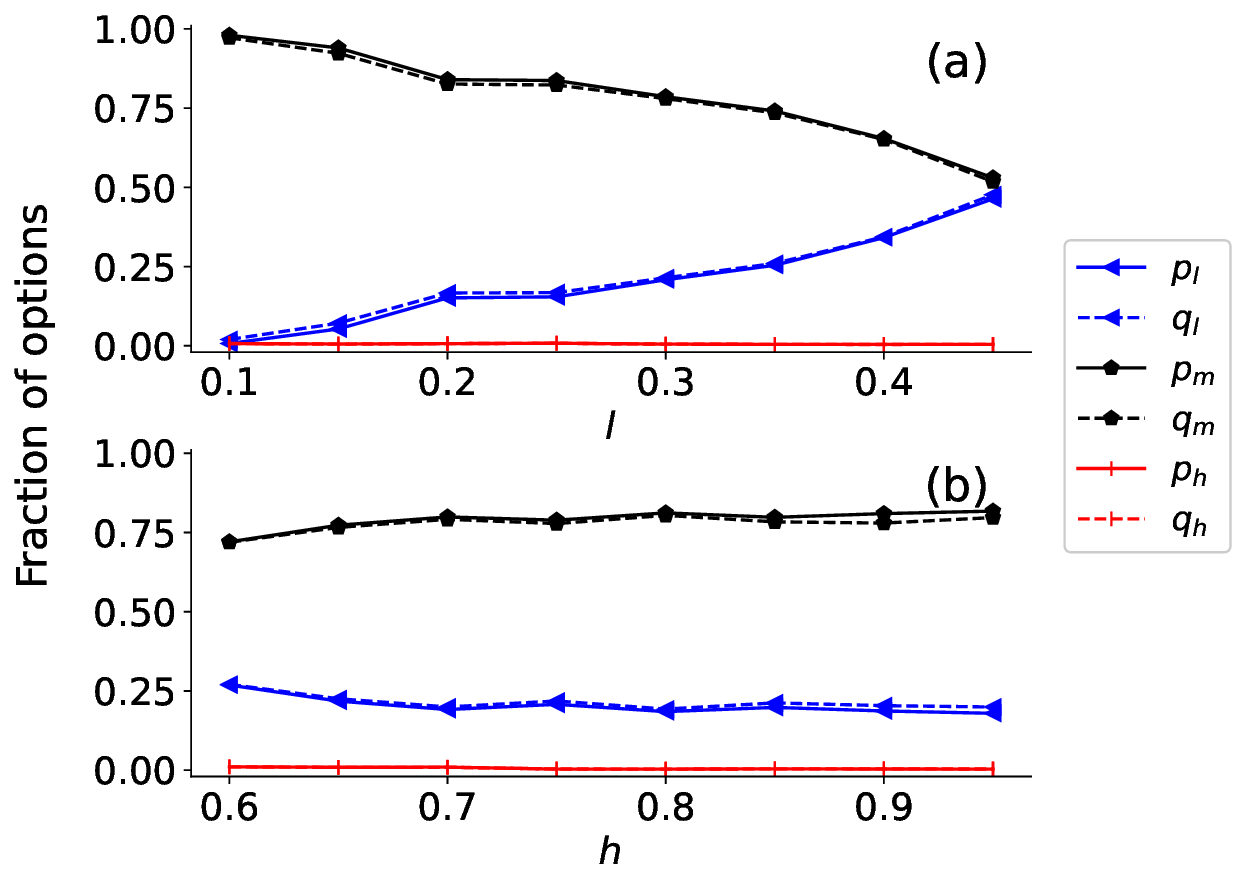}
\caption{Emergence of fairness. 
As in many practices of behavioral experiments, proposers $p$ are offered with three options: mean $(l<0.5)$, fair $(m=0.5)$, and overgenerous $(h>0.5)$, and the responders $q$ have the same acceptance threshold $\{l, m, h\}$. The two subplots show different dependencies of fairness on the l and h. While the rational option fractions $p_l$  and $q_l$ rise as $l$ increases, the marginal impact of $h$ is observed. In all cases, the densities of overgenerous options are always vanishing. Parameters: $\epsilon=0.01$, $\alpha=0.1$, $\gamma=0.9$, $h=0.8$ in (a) and $l=0.3$ in (b).
(Adapted from~\cite{Zheng2025Decoding}).
}
 \label{fig:phaseDigaram_fairness}
 \end{figure}

\section{Fairness} \label{sec:fairness}

Fairness, as a cornerstone of human society, serves as a core norm for resolving conflicts of interest, such as economic inequality, climate justice, and the allocation of public resources. 
To investigate the mechanism of fairness, the ultimatum game (UG)~\cite{Guth1982An} is often adopted as a paradigmatic model. In this game, two players divide an amount of money. One acts as the proposer, who offers  $p$ to the other, called the responder, who has an acceptance threshold $q$. If $p\geq q$, the money is divided as proposed; otherwise, they get nothing.

According to the assumption of \emph{Homo economicus} in the classic Economy, the proposer should make the smallest non-zero offer, as the responder will accept it, since something is better than nothing. Therefore, an extremely unfair outcome is predicted.
However, extensive cross-cultural experiments consistently contradict this prediction, where proposers tend to offer shares around $40\sim50\%$, and approximately 50\% of responders reject unfair offers below 20\% ~\cite{Thaler1988Anomalies,Guth2014More}.

To resolve this fundamental discrepancy, most previous game-theoretic works adopted the imitation learning paradigm~\cite{Szab1998Evolutionary}, where the strategies of better-off peers in the ultimatum game are assumed to spread more readily. 
Within this paradigm, researchers have revealed a bunch of factors, such as spatial structure~\cite{Page2000The,Kuperman2008The,Iranzo2011The}, noise~\cite {Gale1995Learning}, reputation~\cite{Zhang2023Reputation,Deng2025The}, role assignment~\cite{Yang2023Role}, and empathy~\cite{Page2002Empathy} contribute to the emergence of fairness, providing valuable insights into the mechanisms behind its evolution~\cite{Debove2016Models}.

Recently, some studies have turned to the RL paradigm to decode the emergence of fairness~\cite{Zheng2025Decoding,WU2025Q}, where they have shown that the endogenous incentive is sufficient to drive the emergence of fairness and exogenous factors are not needed. 
In Ref.~\cite{Zheng2025Decoding}, they consider a two-player scenario, where the decision-making is empowered by the Q-learning algorithm. The two players take turns to play the role of proposer and responder, and therefore each is associated with two Q-tables to guide each role. The state consists of their strategy combination in the previous round, and three fixed actions for their choices for either role: low (\(p_l,q_l <0.5\)), middle (\(p_m,q_m =0.5\)), or high (\(p_h,q_h >0.5\)) options.
They reveal that when individuals value both historical experience (small \(\alpha\)) and future rewards (large \(\gamma\)), fairness emerges significantly. An important result is that as the offer increases, the successful deals also rise, in line with observations in behavioral experiments, see Fig.~\ref{fig:phaseDigaram_fairness}.
 These results remain robust for different role assignments for the two players, such as rotating, random, fixed role, and for the extended scenario to a $1d$ lattice population.

A different implementation is given in Ref.~\cite{WU2025Q}, where they study the evolution of a 2d spatial UG with a strategy-adjustment Q-learning. Instead of fixed actions, the action set is composed of offer/acceptance threshold increase, decrease, or maintenance, with two sensitivity factors controlling the magnitude of adjustment. They reveal that when the two factors become imbalanced, a promoted fairness is seen. By comparison to the implementation of imitation learning, people empowered by their Q-learning have a higher level of fairness. The study also examined the impact of learning parameters, and they reached a consistent conclusion that the appreciation of historic experience and the future rewards generally yields a high level of fairness.

The mechanism analysis for the emergence of fairness is also conducted in Ref.~\cite{Zheng2025Decoding}, which consists of two phases. In the first stage, those strategies leading to failed deals are removed from the system, and in the second stage, the remaining ones evolve either into the fair strategy \((p_m, q_m)\) or the rational strategy \((p_l,q_l)\) by a branching process. In short, the historical experience enables players to draw lessons from the past, and the expectation in future reward encourages responders to shift from a low offer to pursuing a higher fair offer. This shift may cause some immediate loss, but responders can obtain higher accumulated rewards in the long term by forcing proposers to raise their offer to reach deals.

\section{Resource allocation}\label{sec:Resource}

Resource allocation is a fundamental issue in both nature and human societies, and its efficiency directly affects system stability and sustainability~\cite{Samuelson2005Economics}. Although general equilibrium theory in economics assumes that supply and demand can reach an optimal allocation, how such an optimum is achieved remains unclear. The key question is: how can populations reach an optimal allocation when individuals act in their own self-interest?

The minority game (MG)~\cite{Challet1997Emergence,Challet2005MinorityGI} provides perhaps the simplest toy model for studying this question, inspired by the El Farol bar problem~\cite{Arthur1994Inductive}. Its core logic follows the ``minority-wins” rule: an odd number $N$ of individuals repeatedly choose between two options (e.g., go to the bar or stay home), and those in the minority win. In the seminal work~\cite{Challet1997Emergence}, each agent is assigned a set of predefined, static strategies from a shared pool, and a phase transition is observed as model parameters vary. Many follow-up studies have explored coordination mechanisms~\cite{Challet2005MinorityGI,chakraborti2015statistical}, and various model variants are proposed~\cite{zhou2005self,zhang2013controlling}. A major limitation, however, is that these models rely on fixed strategies and cannot capture real-world adaptive decision-making.

\begin{figure}[tbp]
\centering
\includegraphics[width=0.8\linewidth]{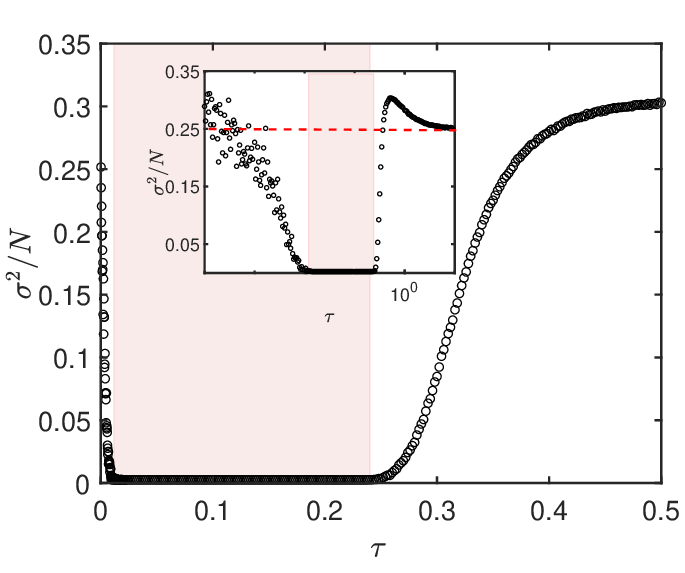}
\caption{Emergence of coordination in the minority game.
The volatility $\sigma^2/N$ (a measure of coordination; lower values indicate better coordination) is plotted as a function of temperature $\tau$, where higher $\tau$ increases the likelihood of random exploration over exploitation. Results are averaged over 100 realizations, with data collected over $5\times10^3$ time steps after a transient period of $5\times10^4$ Monte Carlo steps. The inset shows the same data with a logarithmic $x$-axis, with the shaded region indicating the parameter range where optimal coordination is achieved. The red dashed line at $\sigma^2/N = 0.25$ corresponds to the benchmark case in which players decide randomly (e.g., by coin flip). (Adapted from~\cite{zheng2025optimal}).
}
\label{fig:MG1}
\end{figure}

Recently, RL has brought new vitality and fresh perspectives to this field~\cite{Zhang2024Self,Zhang2019Reinforcement,Zhang2019Collective,Zhang2026Dual,zheng2025optimal,Shao2025Network,Rao2025Emergent}, primarily along two lines: value-based RL (e.g., Q-learning) and policy-based RL (e.g., REINFORCE~\cite{Williams1992Simple}).
Along the value-based line, Q-learning has shown that optimal coordination can emerge under the RL paradigm~\cite{Zhang2026Dual,zheng2025optimal}. Ref.~\cite{zheng2025optimal} studies the MG in the El Farol bar context using Q-learning, where the state is the number of people who went to the bar in the last round, and actions are ``go” or ``not go.” Instead of $\epsilon$-greedy selection, they use a softmax version with a temperature-like parameter to balance exploitation and exploration (Fig.~\ref{fig:MG1}). At low temperatures, the system gets stuck in partially coordinated local optima; at moderate temperatures, optimal coordination emerges when players value both experience and future rewards; at high temperatures, coordination breaks down into anti-coordination, and resource-utilization efficiency drops below the purely random (coin-flip) baseline.
Mechanism analysis reveals there is a symmetry-breaking in action preference, where most people's preferences are stabilized, while one ``pathetic individual" keeps switching, benefiting all others except itself.

In fact, the original MG scheme~\cite{Challet1997Emergence} can be seen as a crude form of RL, where high-scoring strategies are reinforced over time. Given this fact, Ref.~\cite{Zhang2026Dual} explores the synergy between dual RL schemes—some players use classical static strategies, others use Q-learning. The study finds an optimal mixing proportion that maximizes resource allocation, marked by a first-order phase transition. The Q-learning population further self-organizes into internally and externally coordinating clusters. The latter develop momentum strategies similar to those in financial markets, which prevent long-term idling of resources but also yield lower long-term returns for those using them. This work reveals how strategy-level coordination emerges through inter-population heterogeneity.

Note that there are some early attempts~\cite{Zhang2024Self,Zhang2019Reinforcement,Andrecut2001Qlearning} applying RL to solve the MG and claimed that the herding effect is suppressed, but with large persistent fluctuations. The reason lies in their self-regarding setup, where individuals only focus on their own actions and neglect others' choices, leaving them without enough information to coordinate effectively.

Along the second line, the policy-based methods optimize action probabilities directly without estimating a value function, and can also achieve coordination. Ref.~\cite{Rao2025Emergent} introduces a modified REINFORCE algorithm into the MG, using continuous policy probabilities and removing future reward estimation—relying only on historical payoffs. This preserves the ``inductive learning” nature of the original MG, rather than the deductive logic of Q-learning. Their results show a coordination mechanism without symmetry breaking: individual behavior remains nearly stochastic, yet system-wide volatility stays low due to weak anticorrelation at the collective level. This symmetry-preserving coordination is further explored from a network perspective in Ref.~\cite{Shao2025Network}.

\begin{figure*}[htbp]
\centering
\includegraphics[width=0.8\linewidth]{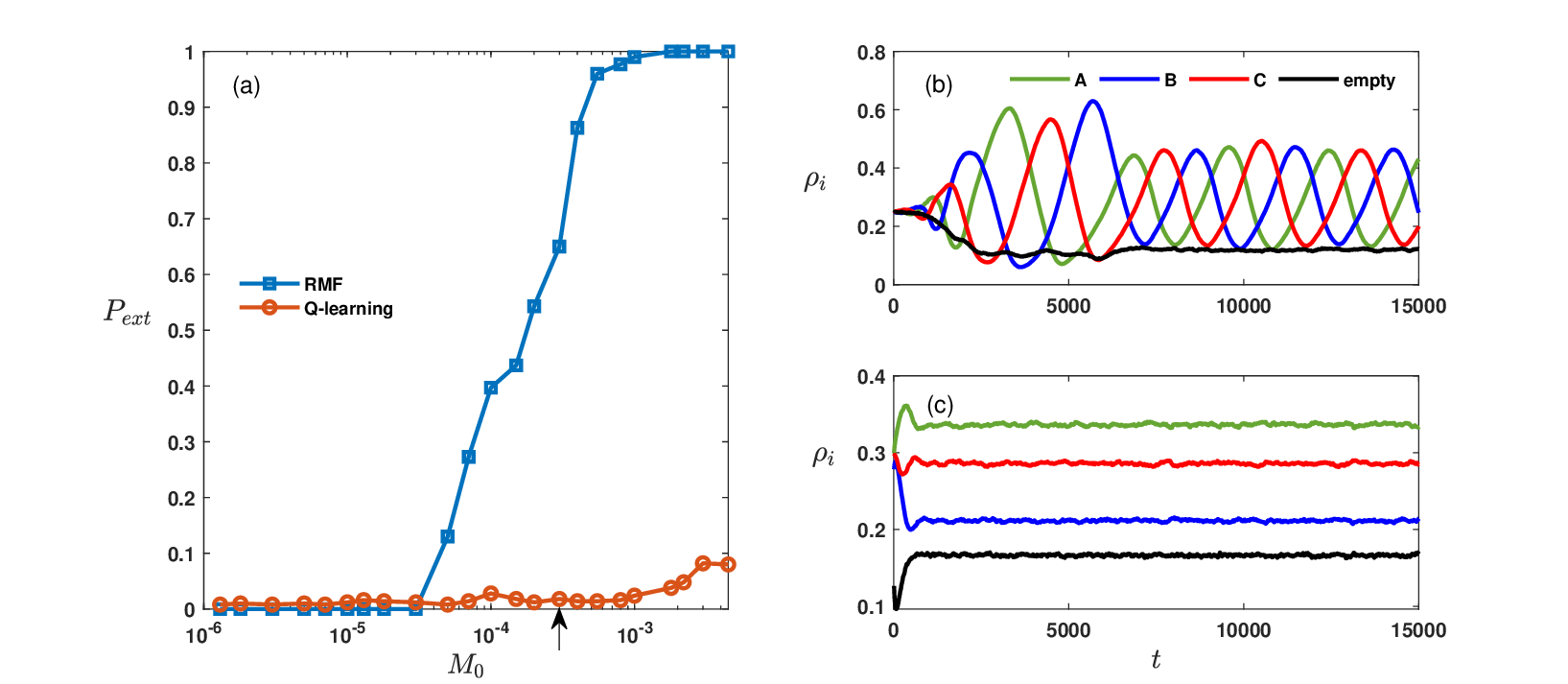}
\caption{Comparison study for species coexistence with the traditional RMF and Q-learning. 
(a) Extinction probability versus the baseline mobility $M_{0}$. The blue squares and red circles represent the results for the traditional RMF model and {the} Q-learning model, respectively. The extinction probability drops significantly after applying the Q-learning algorithm, greatly enhancing system stability. Typical time series of three species densities as well as the density of empty sites for the traditional RMF model (b) and our model (c), both at $M_{0}=3\times10^{-4}$. This means that when species are empowered with reinforcement learning, they are much better at coexisting with each other than the RMF model, and the density oscillation is suppressed. $N=100\times100$. 
(Adapted from~\cite{jiang2025species}).
}
\label{fig:biodiversity}
\end{figure*}

\section{Ecological systems}\label{sec:ecology}

While reinforcement learning has proven effective in decoding several human behaviors, recent studies have extended this paradigm to ecological systems. The underlying rationale is that individuals -- including non-human species -- actively make decisions to better adapt to their environments, which applies to most non-human species.

Much of this research focuses on predator-prey systems~\cite{2015Co-evolution,2019Deep-Reinforcement,2020Wang,2021Co-Evolution}, where Q-learning is employed to examine how individuals enhance their survival through learning. These works reveal that while predator learning tends to stabilize ecosystems, prey learning often induces oscillations in species densities -- sometimes even triggering system collapse. A recent study~\cite{2023Predator-prey} further demonstrates that survival pressure in RL-driven predator -- prey systems can lead to the emergence of swarming behavior. Similarly, a flocking model based on neighbor-loss minimization~\cite{durve2020learning} reproduces polarized swarming akin to phenomena observed in the Vicsek model.

Another research direction addresses biodiversity -- a central theme in theoretical ecology. The rock-paper-scissors (RPS) game serves as a canonical model of cyclic dominance, yet in spatial settings, coexistence is not guaranteed. In the seminal spatial RPS model by Reichenbach et al.~\cite{2007Mobility}, three species reproduce, predate, and migrate on a two-dimensional domain, usually called the RMF model. They coexist via spiral waves at low mobility, but go extinct when mobility exceeds a threshold — a prediction at odds with real-world observations of highly mobile species coexisting in nature.

To resolve this discrepancy, a recent spatial RPS model incorporates a joint Q-learning algorithm~\cite{jiang2025species}, where individuals of the same species share a common Q-table, updated collectively over generations—mirroring natural collective intelligence. Rewards are tied to survival and predation success. This framework shows that with Q-learning-guided migration, extinction becomes rare even under high mobility (Fig.~\ref{fig:biodiversity}(a)). Analysis reveals that individuals develop two key tendencies: escaping predators and staying near prey. These behavioral heterogeneities suppress spiral wave formation and damp density oscillations, thereby promoting coexistence (Fig.~\ref{fig:biodiversity}(b,c)). However, the imbalance between these tendencies undermines behavioral heterogeneity and jeopardizes stable coexistence.

Beyond these lines of inquiry, deep RL has also been applied to collaborative hunting~\cite{2024Collaborative}, revealing that sophisticated coordination can emerge without high-level cognition. Other studies integrate RL with realistic ecosystems to reproduce empirical behavioral and population patterns~\cite{strannegaard2025predicting}, paving the way for predicting ecosystem resilience and tipping points. Additional work demonstrates RL-enabled path planning in complex settings~\cite{nasiri2022reinforcement}, and successful navigation for microswimmers in noisy environments~\cite{muinos2021reinforcement}.
Together, these efforts underscore the significant potential of RL — both theoretical and experimental — in deciphering and predicting ecological dynamics.

\section{Concluding remarks}\label{sec:summary} 

{Compared with mainstream imitation learning paradigms, reinforcement learning offers a fundamentally different and introspective approach to understanding behavior — one that provides a novel perspective on the origins of many human traits. Central to reinforcement learning is its long-term orientation: individuals make decisions based on cumulative future rewards rather than immediate gains.}

Within this framework, {the above work demonstrates} how four key human behaviors — cooperation, trust, fairness, and optimal resource allocation — can be explained in a unified manner. These emerge naturally when players value experience and maintain long-term objectives, without relying on external assumptions. Beyond human behaviors, the RL paradigm is also applicable to ecological systems, illustrating how it accounts for species coexistence.


{Collectively, these achievements position RL as a potential unifying theoretical framework applicable to a variety of complex systems where agents—whether human or non-human—are capable of making active decisions. Moreover, RL introduces novel elements such as Q-tables, offering a unique lens through which to understand the psychological evolutionary processes underlying behavior. More critically, the value of RL extends beyond iteratively optimizing existing strategies based on cumulative rewards; its capacity to autonomously explore and discover optimal strategies that surpass human-prescribed designs further expands the boundaries of this paradigm~\cite{su2025multi}. This endows RL with a dual function—both updating existing strategies and discovering new ones—thereby providing richer research perspectives and methodological support for subsequent analyses of behavioral evolution in various complex systems. It's important to note that RL is not a superior replacement of IL; instead, the two paradigms are complementary and are a context-dependent choice for the system under study.}

{However, this promising picture should not obscure the fact that several fundamental challenges remain. As research progresses, a series of key questions warranting further exploration has come to the fore, particularly regarding the realism of state representations and the experimental validation of reinforcement learning paradigms in real-world human decision-making.}

{First, realistically portraying state representation constitutes a core challenge—how to more authentically reconstruct the information environment in which individuals make decisions within models remains an inadequately addressed issue. Most current research pays insufficient attention to the cognitive foundations and informational constraints underlying state settings. In real-world scenarios, individuals often face objective conditions such as incomplete information, limited perceptual capabilities, and cognitive load constraints. Furthermore, significant heterogeneity exists across individuals in terms of information access channels and processing methods. State design is thus perpetually confronted with a dilemma: incorporating excessive information can easily lead to dimensional explosion, substantially increasing learning complexity; whereas overly simplified information may restrict an individual's ability to explore and acquire optimal strategies. Consequently, investigating how to achieve effective dimensionality reduction of information and how to systematically integrate more realistic cognitive constraints into the modeling process will provide crucial support for narrowing the gap between theoretical models and real-world decision-making scenarios.}

{Second, although RL has demonstrated immense potential as a unified theoretical paradigm, its core foundational principles still lack sufficient direct validation through behavioral experiments—a fact that significantly impedes the development of a robust theoretical framework. It is worth noting that existing relevant behavioral experimental research~\cite{Wang2017Onymity,Wang2018Exploiting,Wang2020Communicating,Jia2025Social} has already accumulated rich experimental paradigms and empirical evidence. These data can serve as resources for subsequent validation of the foundational principles of RL and optimization of state representations.}

{Third, while RL has demonstrated success in replicating experimental observations, it is important to note that these explanatory capabilities are not unique to this paradigm; other bounded rationality models, and adaptive learning models may likewise reproduce similar behavioral patterns. Therefore, a systematic comparison between reinforcement learning and alternative learning rules—evaluated from the dual perspectives of experimental fit and theoretical predictive power—represents an important and promising direction for future research.}
 
\section*{Acknowledgments}
We would like to thank Professor Weiran Cai and all other collaborators who have contributed to the research on this topic. 
This work is supported by the National Natural Science Foundation of China (Grants Nos. 12075144, 12165014), the Fundamental Research Funds for the Central Universities (Grant No. GK202401002), and the Key Research and Development Program of Ningxia Province in China (Grant No. 2021BEB04032).

\end{CJK*}
\end{document}